\documentclass[10pt]{artikel3}
\usepackage{amsfonts,amsmath,amssymb}
\usepackage{pslatex}
\usepackage{float,braket,color}
\usepackage{subfigure,hyperref}
\usepackage{graphicx,wrapfig}

\newcommand{\p}{\partial}
\newcommand{\oc}{\overline{c}}

\renewcommand{\d}{\ensuremath{\mathrm{d}}}

\renewcommand{\d}{\ensuremath{\mathrm{d}}}

\newcommand{\YM}{\ensuremath{\mathrm{YM}}}
\newcommand{\e}{\ensuremath{\mathrm{e}}}
\newcommand{\GZ}{\ensuremath{\mathrm{GZ}}}
\usepackage[small,bf]{caption}
\newcommand{\sect}[1]{ \section{#1} \setcounter{equation}{0} }

\definecolor{NeleBlue}{rgb}{.255,.41,.884} 
\definecolor{NeleRed}{rgb}{1, 0, 0} 
\definecolor{NeleGreen}{rgb}{.196,.804,.196} 
\definecolor{NeleYellow}{rgb}{1,.648,0} 

\begin{document}
\title{{\bf On the reanimation of a local BRST invariance in the (Refined) Gribov-Zwanziger formalism}}
\author{D.~Dudal\thanks{david.dudal@ugent.be} ~\& N.~Vandersickel\thanks{nele.vandersickel@ugent.be}\\\\
\small \textnormal{Ghent University, Department of Physics and Astronomy, Krijgslaan 281-S9, 9000 Gent, Belgium}\normalsize}

\date{}
\maketitle
\begin{abstract}
We localize a previously established nonlocal BRST invariance of the Gribov-Zwanziger (GZ) action by the introduction of additional fields. We obtain a modified GZ action with a corresponding local, albeit not nilpotent, BRST invariance. We show that correlation functions of the original elementary GZ fields do not change upon evaluation with the modified partition function. We discuss that for vanishing Gribov mass, we are brought back to the original Yang-Mills theory with standard BRST invariance.
\end{abstract}

\sect{Introduction}
In any textbook on quantum field theory, one can find the well-established Faddeev-Popov (FP) procedure \cite{Faddeev:1967fc}, which allowed for a clean quantization of gauge theories. In particular, choosing the Landau gauge, $\p_\mu A_\mu=0$, the gauge fixed partition function for an Euclidean SU($N$) Yang-Mills gauge theory reads
\begin{eqnarray}\label{inl1}
Z=\int [\d\Phi] \delta\left[\p_\mu A_\mu^a\right] \det\left[-\p_\mu D_\mu^{ab}\right] \e^{-S_{YM}}\,,\qquad \text{with } S_{YM}=\int \d^d x \frac{1}{4}F_{\mu\nu}^2\,.
\end{eqnarray}
Throughout this letter, $\int[\d\Phi]$ will denote path integration over all fields occurring in the exponentiated action. $Z$ can be equivalently expressed with a manifestly local action,
\begin{eqnarray}\label{inl2}
Z=\int [\d\Phi] \e^{-S_{YM}+S_{GF}}\,,\qquad \text{with } S_{GF}=\int \d^d x \left(b^a\p_\mu A_\mu^a + \overline c^a \p_\mu D_\mu^{ab}c^b\right)\,,
\end{eqnarray}
by the introduction of the (anti-)ghost fields $\overline c^a$ and $c^a$. With $\mathcal{M}^{ab}=-\p_\mu D_\mu^{ab}$, where $D_\mu^{ab}=\p_\mu\delta^{ab}-gf^{abc}A_\mu^c$, we denote the FP operator, whose determinant corresponds to the Jacobian accompanying the $\delta$-function. During this derivation, it was tacitly assumed that $\p_\mu A_\mu^a=0$ only displays a single root along the gauge orbit of any given $A_\mu$.

Later on, it was found that this action enjoys the BRST symmetry \cite{Becchi:1975nq}, generated by
\begin{eqnarray}\label{GZb3a}
sA_{\mu }^{a} ~=~-D_{\mu }^{ab}c^b\,,\quad  sc^{a} ~=~\frac{1}{2}gf^{abc}c^{b}c^{c}\,,\quad s\overline{c}^{a} ~=~b^{a}\,, \quad  sb^{a}~=~0\,.
\end{eqnarray}
We recall that this generator is nilpotent, $s^2=0$. A somewhat more general setting is the BRST quantization \cite{Baulieu:1983tg,Henneaux:1992ig}, in which case the gauge is fixed by adding a physically irrelevant BRST exact term to the classical action, in analogy with $S_{FP}= s \int \d^dx\left(\overline c^a \p_\mu A_\mu^a\right)$. Anyhow, whatever method of quantization one is using, the BRST symmetry is always encountered. It plays a pivotal role in proving the (perturbative) unitarity and renormalizability of gauge theories to all orders, see e.g.~\cite{Becchi:1975nq,Baulieu:1983tg,Henneaux:1992ig,Kugo:1979gm,Barnich:2000zw,Piguet:1995er}.

The foregoing results are however only valid at the perturbative level, see e.g.~\cite{Piguet:1995er,becchi3}. If we leave the high energy region, i.e.~the energy sector where asymptotic freedom guarantees a small coupling and thus validates a perturbative treatment, things get more obscure. In particular, one encounters the problem of Gribov copies \cite{Gribov:1977wm}: there can be, and there are, multiple solutions to the desired gauge condition. The latter observation is at odds with the assumptions made in the FP quantization procedure. Some gauges can escape this problem, although usually these gauges are then suffering from other problems in the continuum like a lack of covariance, problems with locality, boundary conditions, $\ldots$.

To our knowledge, there is no waterproof way of dealing with the issue of Gribov copies. The most renown attempt is based on Gribov's seminal approach \cite{Gribov:1977wm}, later worked out to all orders in an alternative way by Zwanziger in a series of papers, see e.~g.~\cite{Zwanziger:1989mf,Zwanziger:1992qr,Zwanziger:1993dh} and references therein. Let us refer to \cite{Sobreiro:2005ec,Zwanziger:2010iz} for an overview of the underlying ideas and historical references. Crudely summarizing, it was observed that zero modes $\omega$ of the FP operator, $\p_\mu D_\mu \omega=0$, correspond to (infinitesimal) gauge copies of Landau gauge configurations, since $\p_\mu A_\mu=\p_\mu A_\mu'$ for $A_\mu'=A_\mu+D_\mu\omega$, hence a natural requirement would be to restrict the integration region in $A_\mu$-space to those with a positive FP operator. Upon some assumptions and simplifications, Zwanziger was able to construct the following partition function
\begin{eqnarray}\label{GZb1}
Z=\int [d\Phi] \e^{-S_{GZ}}\,,\quad S_{GZ} &=& S_{YM} + S_{GF}   + \int \d^d x\left( \overline \varphi_\mu^{ac} \p_\nu D_\nu^{ab} \varphi_\mu^{bc}  - \overline \omega_\mu^{ac} \p_\nu D_\nu^{ab} \omega_\mu^{bc}  +g\gamma ^{2}  f^{abc}A_\mu^a \left( \varphi_\mu^{bc} +  \overline \varphi_\mu^{bc}\right) \right.\nonumber\\&&\left.- g f^{abc} \p_\mu \overline \omega_\nu^{ae}    D_\mu^{bd} c^d  \varphi_\nu^{ce}+\,\gamma^4 d (N^2 - 1 )\right)\,,
\end{eqnarray}
which would implement, in the continuum, the restriction to the Gribov region $\Omega=\left\{A_\mu\vert \p_\mu A_\mu=0\,, -\p_\mu D_\mu>0\right\}$. The fields $\{\varphi_\mu^{bc},\overline\varphi_\mu^{bc}\}$ are bosonic, while the $\{\omega_\mu^{bc},\overline\omega_\mu^{bc}\}$ are Grassmannian. The parameter $\gamma$ has the dimension of mass, and it is not free, but must be self-consistently fixed as the nonzero solution of the horizon condition, $\frac{\p \Gamma}{\p \gamma^2}=0$, if $Z=\e^{-\Gamma(\gamma^2)}$, to restore the connection with the original gauge theory \cite{Zwanziger:1989mf,Zwanziger:1992qr}. It can be proven that \eqref{GZb1} constitutes a renormalizable quantum field theory \cite{Zwanziger:1992qr,Maggiore:1993wq,Dudal:2010fq}.

This restriction is not necessarily the end of the story, as $\Omega$ still contains a subset of copies \cite{vanBaal:1991zw}, apparently necessitating a further restriction to the so-called Fundamental Modular Region $\Lambda\subset\Omega$, although some argument was given that at the level of correlation functions, restricting to $\Omega$ or $\Lambda$ would be equivalent \cite{Zwanziger:2003cf}.

The BRST transformation \eqref{GZb3a} is naturally extended to the fields as\footnote{In this fashion, they are BRST doublets when $\gamma=0$, in which case we have complete equivalence between $S_{GZ}$ and $S_{YM}+S_{GF}$.}
\begin{eqnarray}\label{GZb3b}
s\varphi _{\mu}^{ac} ~=~\omega_{\mu}^{ac}\,,\quad s\omega_{\mu}^{ac}~=~0\,,\quad s\overline{\omega}_{\mu}^{ac} ~=~\overline{\varphi }_{\mu}^{ac}\,,\quad s \overline{\varphi }_{\mu}^{ac}~=0~\,.
\end{eqnarray}
It is now easily checked that
\begin{eqnarray}\label{GZb3bbis}
s S_{GZ}&=& g\gamma^2\int\d^d x\left(f^{abc}A_\mu^a \omega_\mu^{bc}-f^{abc}(\varphi_\mu^{bc}+\overline\varphi_\mu^{bc})D_\mu^{ak}c^k\right)\,,
\end{eqnarray}
so we encounter a soft\footnote{With soft, we mean that it is proportional to a mass parameter, so it can be controlled at the quantum level. In the recent paper \cite{Capri:2010hb}, it was even shown that the BRST breaking can be transformed into a linear breaking by additional sets of auxiliary fields. This provides an efficient and powerful way to (re)discuss (non)renormalization issues of the Gribov-Zwanziger formalism, based on the cohomology of the adapted BRST. If we would use the techniques of this paper to construct a local symmetry, thus not linearly-broken, for the action proposed in \cite{Capri:2010hb}, it turns out that almost the same can be done as shall be presented here, leading to a modified version of the action of \cite{Capri:2010hb}, including the original sets of fields of \cite{Capri:2010hb}, supplemented with the ones we shall introduce in this work.} breaking of the usual BRST $s$. This breaking is of a nonperturbative nature, as $sS_{GZ}\sim \gamma^2\sim \Lambda_{QCD}^2$, the latter due to the horizon condition. On general grounds it was argued in \cite{Dudal:2008sp} that one should not hope to recover a locally modified BRST transformation of the action $S_{GZ}$.

A potential alternative to the Gribov-Zwanziger approach would be to not just select a single gauge representative, but to try to average over all gauge copies in some manner, and try to look at the BRST in this different setup, as proposed in \cite{vonSmekal:2007ns,vonSmekal:2008es}. So far, this was however only explicitly studied for low-dimensional compact QED, where analytical results are possible.  The (numerical) extension to more dimensions, or to the non-Abelian version of this construction, have, to our knowledge, not yet been presented explicitly. For other lattice inspired work about the BRST in connection to Gribov copies, see \cite{Burgio:2009xp}.

Although not everybody is convinced about the physical relevance, if any, of Gribov copies, one cannot deny that they are present in the Landau gauge, be it numerically or analytically \cite{vanBaal:1991zw,Cucchieri:1997dx,Greensite:2010hn,Maas:2009ph}, and that taking them into account seriously influences the results for e.g.~the gluon and ghost propagators in the infrared sector. Although these propagators themselves are not physical, they are frequently linked to e.g.~confinement criteria as in \cite{Braun:2007bx}. We recall that confinement is still one of the most intriguing and yet not fully understood problems in QCD \cite{Alkofer:2006fu}. These propagators have witnessed a great deal of interest in recent years, both from numerical or analytical viewpoint \cite{Zwanziger:2010iz,Zwanziger:2003cf,Dudal:2008sp,vonSmekal:2007ns,Alkofer:2000wg,Dudal:2007cw,Maas:2009se,Cucchieri:2007md,Aguilar:2008xm,Binosi:2009qm,Fischer:2008uz,Bogolubsky:2009dc,Sternbeck:2005tk,Bornyakov:2009ug,Boucaud:2008ky,Gracey:2005cx,Huber:2009tx,Oliveira:2008uf,Aguilar:2004sw,Dudal:2010tf,Furui:2005bu,Oliveira:2010xc,Bornyakov:2009ug}. Although one does not always rely on the formulation \eqref{GZb1} when using functional methods as in Schwinger-Dyson studies, with the exception of \cite{Huber:2009tx}, the Gribov ambiguity is at least partially dealt with in an indirect way by imposing suitable boundary conditions \cite{Fischer:2008uz,Binosi:2009qm}.

The unraveled situation looks a bit unpleasant. We recall that in ordinary Yang-Mills gauge theories, the nilpotent BRST symmetry plays a pivotal role in the construction of a physical subspace, viz.~in the identification of relevant operators, through its cohomology \cite{Henneaux:1992ig,Kugo:1979gm,Barnich:2000zw,Piguet:1995er}. To assure that the subspace remains intact after time evolution/interaction, one usually invokes a symmetry to define such subspace. As is well known, at the perturbative level, the BRST symmetry can be used to define a physical subspace of two transverse gluon polarizations, while the 2 remaining unphysical gluon polarizations cancel with the (anti)ghosts \cite{Henneaux:1992ig,Kugo:1979gm}. In general, the solution of the quantum BRST cohomology is given by the classically gauge invariant operators, which mix with physically trivial BRST exact operators and equation of motion related terms \cite{Barnich:2000zw,Piguet:1995er}. In a confining theory as Yang-Mills gauge theories, the physical excitations corresponding to these composite operators should correspond to colorless bound states of gluons, i.e.~glueballs. As the simplest example, we can mention the scalar glueball, which ought to be described by $F_{\mu\nu}^2$. In the GZ theory, the gluon degrees of freedom are believed to be unphysical due to a violation of positivity, already prominent at tree level, and in this sense one can interpret the gluons as being ``confined'', see e.g.~\cite{Sorella:2010fs} and the introduction of \cite{Sorella:2010it} and references therein. The question then arises what the physical degrees of freedom would be? One expects that exactly these would be the glueballs. However, the absence of a (BRST-like) symmetry prevents one from defining a, potentially physical, subspace of suitable quantum extensions of the classically gauge invariant operators like $F_{\mu\nu}^2$. The softly broken BRST $s$ of \eqref{GZb3a}-\eqref{GZb3b} can nevertheless still be used to construct a renormalizable scalar glueball operator $F_{\mu\nu}^2$ \cite{Dudal:2009zh}. However, renormalizability is not the only requirement for a decent quantum operator. In order to be physical, in the sense that its two-point function corresponds to the propagation of a physical degree of freedom, that two-point function must obey certain analyticity properties, see e.g.~\cite{Sorella:2010it,Peskin:1995ev,Baulieu:2009ha} for a short overview. At lowest order however, the operator $F_{\mu\nu}^2$ will lead to unwanted analyticity properties, as discussed in \cite{Zwanziger:1989mf,Baulieu:2009ha}. The concept of an invariant subspace, defined by a symmetry constraint, should then establish a unitary theory of glueballs. Unfortunately, till now, we are lacking any form of BRST like symmetry of the GZ action.

In this letter, we pursue to offer a possible way to overcome this problem by explicitly constructing a modified but nonetheless equivalent version of the Gribov-Zwanziger action $S_{GZ}$ of \eqref{GZb1}, which exhibits a locally modified version of the standard BRST symmetry, though it will turn out to be not nilpotent. We also show that the Refined Gribov-Zwanziger formalism \cite{Dudal:2007cw,Dudal:2008sp} fits with this modified BRST.

\sect{Preliminaries}
We start from the standard GZ action \eqref{GZb1}, set $g\gamma^2=\theta^2$ and drop the vacuum term for notational shortness\footnote{Obviously, this vacuum term will not influence any variation of the action.},
\begin{eqnarray}\label{GZb2}
 S_{GZ} &=& S_\YM +  \int \d^d x\,\left( b^a \p_\mu A_\mu^a +\overline c^a \p_\mu D_\mu^{ab} c^b \right) \nonumber \\&&+ \int \d^d x\left( \overline \varphi_\mu^{ac} \p_\nu D_\nu^{ab} \varphi_\mu^{bc}  - \overline \omega_\mu^{ac} \p_\nu D_\nu^{ab} \omega_\mu^{bc}  +\theta^{2}  f^{abc}A_\mu^a \left( \varphi_\mu^{bc} +  \overline \varphi_\mu^{bc}\right) \underline{- g f^{abc} \p_\mu \overline \omega_\nu^{ae}    D_\mu^{bd} c^d  \varphi_\nu^{ce}} )\right)\,.
\end{eqnarray}
We shall first base ourselves on the reasoning made in \cite{Sorella:2009vt} to identify a nonlocal BRST invariance. Therefore, following \cite{Sorella:2009vt}, we shall for the moment also drop the underlined term. Dropping this term temporarily only leads to a breaking in the BRST $s$ which is itself the $s$-variation of something, thus it is rather harmless. Later on, we shall take this term into account anyhow by the nature of the construction itself. For now, we shall thus study the following GZ action,
\begin{eqnarray}\label{GZb4}
\hat S_{\GZ} &=& S_\YM +  \int \d^d x\,\left( b^a \p_\mu A_\mu^a +\overline c^a \p_\mu D_\mu^{ab} c^b \right)  +\nonumber\\&&+ \int \d^d x\left( \overline \varphi_\mu^{ac} \p_\nu D_\nu^{ab} \varphi_\mu^{bc}  - \overline \omega_\mu^{ac} \p_\nu D_\nu^{ab} \omega_\mu^{bc}  +\theta^{2}  f^{abc}A_\mu^a \left( \varphi_\mu^{bc} +  \overline \varphi_\mu^{bc}\right)\right)\,.
\end{eqnarray}
Applying \eqref{GZb3a} and \eqref{GZb3b} yields
\begin{eqnarray}\label{GZb5}
    s \hat S_{\GZ}=\int \d^d x\left(\theta^2c^k D_\mu^{ka}f^{abc}\left(\varphi_\mu^{bc} +  \overline \varphi_\mu^{bc}\right)+\theta^2f^{abc}A_\mu^a\omega_\mu^{bc}
    +\underbrace{gf^{abc}(D_\nu^{bp}c^p)\left(\p_\nu\overline\varphi_\mu^{ae}\varphi_\mu^{ce}-\p_\nu\overline\omega_\mu^{ae}\omega_\mu^{ce}\right)}_{s(g f^{abc} \p_\mu \overline \omega_\nu^{ae}    D_\mu^{bd} c^d  \varphi_\nu^{ce})}\right)\,.
\end{eqnarray}
According to \cite{Sorella:2009vt}, the positivity of the FP operator inside the Gribov region allows to rewrite
\eqref{GZb4} as
\begin{eqnarray}\label{GZb6}
s\hat S_{\GZ}  &=& \int \d^d x\left( c^a D_\nu^{ab}\Lambda_\nu^b + \theta^2 f^{abc} A_\mu^a \omega_\mu^{bc}\right)\nonumber\\
&=&\int \d^d x\left((D_\nu^{ma}\Lambda_\nu^a)[(\p_\nu D_\nu)^{-1}]^{mc}\frac{\delta}{\delta \overline c^c}S_{\GZ}-\theta^2f^{abc}A_\mu^a[(\p_\nu D_\nu)^{-1}]^{bm}\frac{\delta}{\delta \overline\omega_\mu^{mc}}S_{\GZ}\right)\,,
\end{eqnarray}
with
\begin{equation}\label{GZb8}
    \Lambda_\nu^a= \theta^2 f^{abc}(\varphi_\nu^{bc}+\overline\varphi_\nu^{bc})-gf^{bap}\left(\p_\nu\overline\varphi_\mu^{bc}\varphi_\mu^{pc}-\p_\nu\overline\omega_\mu^{bc}\omega_\mu^{pc}\right)\,.
\end{equation}
From \eqref{GZb6}, we can read off a new, albeit nonlocal, BRST symmetry, $s'\hat S_{\GZ}=0$, generated by
\begin{eqnarray}\label{GZb7}
s'A_{\mu }^{a} &=&-D_{\mu }^{ab} c^b\,, \quad s'c^{a} ~=~\frac{1}{2}gf^{abc}c^{b}c^{c}\,, \quad s'\overline{c}^{a} ~=~b^{a}-(D_\nu^{kc}\Lambda_\nu^c)[(\p_\nu D_\nu)^{-1}]^{ka}\,, \quad   s'b^{a}~=~0\,,  \nonumber \\
s'\varphi _{\mu}^{ac} &=&\omega_{\mu}^{ac}\,,\quad s'\omega_{\mu}^{ac}~=~0\,, \quad s'\overline{\omega}_{\mu}^{ac} ~=~\overline{\varphi }_{\mu}^{ac}+\theta^2f^{qpc}A_\mu^q[(\p_\nu D_\nu)^{-1}]^{pa}\,,\quad s' \overline{\varphi }_{\mu}^{ac}~=~0\,.
\end{eqnarray}
We draw attention to the fact that $s'$ is not nilpotent, $s'^2\neq0$, \cite{Sorella:2009vt}.

\sect{Localization of the BRST variations}
We want to explore the possibility to localize the nonlocal expressions appearing in the BRST variations \eqref{GZb7}. We have in mind to introduce extra fields into the GZ action, in such a way that their equation of motions reproduce the nonlocal BRST expressions. As such, we can hope to establish a (at least on-shell) local version of the BRST symmetry $s'$. In this context, we wish to mention that in \cite{Kondo:2009qz}, a non-local nilpotent BRST symmetry was constructed. Unfortunately, due the rather complicated structure of the result of \cite{Kondo:2009qz}, we have not been able to find a local version of that BRST. As it shall soon become clear, our localization procedure starts from the local GZ action itself, and at the end, we shall naturally come to the non-local BRST just described upon using some equations of motion. The non-local symmetry of \cite{Kondo:2009qz} seems to fall outside this construction.

We shall treat the breaking proportional to $\Lambda_\nu^a$ in two parts and we introduce the notation
\begin{equation}\label{GZb8bis}
    \overline\Lambda_\nu^a= f^{abc}(\varphi_\nu^{bc}+\overline\varphi_\nu^{bc})\,,\qquad \hat{\Lambda}_\nu^a=-gf^{bap}\left(\p_\nu\overline\varphi_\mu^{bc}\varphi_\mu^{pc}-\p_\nu\overline\omega_\mu^{bc}\omega_\mu^{pc}\right)
\end{equation}
for the true, resp.~ fake BRST breaking content of $\Lambda_\nu^a\equiv\theta^2 \overline\Lambda_\nu^a+\hat \Lambda_\nu^a$ . The reason for this is that it will naturally lead to a modification of the \emph{complete} GZ action \eqref{GZb1} rather than of the reduced version \eqref{GZb4}.

\subsection{Auxiliary action}
We start with the original BRST $s$, and introduce the following doublets
\begin{eqnarray}\label{GZb9}
s \alpha^a &=&\Omega^a\,,\quad   s\Omega^a~=~0\,,\quad s \overline\Omega^a ~=~\overline\alpha^a\,,\quad   s\overline\alpha^a~=~0\,,\nonumber\\
s \beta_\mu^{ab} &=&\Psi_\mu^{ab}\,, \quad  s\Psi_\mu^{ab}~=~0\,,\quad s \overline\Psi_\mu^{ab} ~=~\overline\beta_\mu^{ab}\,,\quad  s\overline\beta_\mu^{ab}~=0~\,,
\end{eqnarray}
The $\alpha^a$, $\overline{\alpha}^a$, $\beta_\mu^{ab}$ and $\overline\beta_\mu^{ab}$ are commuting, while $\Omega^a$, $\overline{\Omega}^a$, $\Psi_\mu^{ab}$ and $\overline\Psi_\mu^{ab}$ are anti-commuting fields. We also introduce the auxiliary action
\begin{eqnarray}\label{GZb11}
S_{aux}&=&s\int \d^d x\left(\alpha^a \p_\mu D_\mu^{ab}\overline\Omega^b -\overline\Omega^a D_\nu^{ab} \overline\Lambda_\nu^b+ \beta_\nu^{ac}\p_\mu D_\mu^{ab}\overline{\Psi}_\nu^{bc}-f^{abc}A_\mu^a\overline{\Psi}_\mu^{bc}\right)\nonumber\\
&=&\int \d^d x\left(\Omega^a \p_\mu D_\mu^{ab}\overline \Omega^b + \alpha^a \p_\mu D_\mu^{ab}\overline \alpha^b + gf^{abc}(\p_\mu \alpha^a)(D_\mu^{bd}c^d)\overline \Omega^c-\overline\alpha^a D_\nu^{ab} \overline\Lambda_\nu^b+\overline\Omega^a s(D_\nu^{ab} \overline\Lambda_\nu^b)\right.\nonumber\\
&&\left.+ \Psi_\nu^{ac}\p_\mu D_\mu^{ab}\overline \Psi_\nu^{bc}+\beta_\nu^{ac}\p_\mu D_\mu^{ab}\overline \beta_\nu^{bc} +gf^{abc} (\p_\mu\beta_{\nu}^{ae})(D_\mu^{bd}c^d)\overline \Psi_\nu^{ce}-f^{abc}A_\mu^a\overline{\beta}_\mu^{bc}-f^{abc}\overline{\Psi}_\mu^{bc}D_\mu^{ad}c^d\right)\,.
\end{eqnarray}
It is clear at sight that the equations of motions for $\alpha^a$ and $\beta_\mu^{ab}$ are closely related to the $\theta$-dependent part of the nonlocal expressions in the r.h.s.~of \eqref{GZb7}.

For the moment, let us just change the GZ action \eqref{GZb4} by hand and consider
\begin{eqnarray}\label{GZb13}
\hat S_{\GZ}^{mod}=\hat S_{\GZ}+S_{aux}\,,
\end{eqnarray}
and define the transformation\footnote{$\delta$ itself will not correspond to a symmetry.} $\delta$ by means of
\begin{eqnarray}\label{GZb14}
    \delta \overline\alpha^a &=&\theta^2 c^a\,,   \delta \overline{c}^a~=~-\theta^2 \alpha^a\,, \delta \overline\beta_\mu^{bc} ~=~\theta^2 \omega_\mu^{bc}\,,     \delta \overline{\omega}_\mu^{bc}~=~\theta^2\beta_\mu^{bc}\,,\delta(\text{rest})~=~0\,.
\end{eqnarray}
Then, we find
\begin{eqnarray}\label{GZb15}
    (s+\delta)(\hat S_{\GZ}^{mod})&=&  \underbrace{s\hat S_{\GZ}}_{*} + \underbrace{\delta \hat S_{\GZ}}_{**} + \underbrace{\delta S_{aux}}_{***}\nonumber\\
    &=&\underbrace{\int \d^d x\left(\theta^2 c^a D_\nu^{ab}\overline\Lambda_\nu^b + \theta^2 f^{abc} A_\mu^a \omega_\mu^{bc}+s(g f^{abc} \p_\mu \overline \omega_\nu^{ae}    D_\mu^{bd} c^d  \varphi_\nu^{ce})\right)}_{*}\nonumber\\
    &&+\underbrace{\int \d^dx \left(-\theta^2 \alpha^a \p_\mu D_\mu^{ab}c^b-\theta^2\beta_\mu^{ac}\p_\nu D_\nu^{ab}\omega_\mu^{bc}\right)}_{**}\nonumber\\
    &&+\underbrace{\int \d^dx \left(\theta^2 \alpha^a \p_\mu D_\mu^{ab}c^b -\theta^2c^a D_\nu^{ab}\overline\Lambda_\nu^b+\theta^2 \beta_\mu^{ac}\p_\nu D_\nu^{ab}\omega_\nu^{bc}-\theta^2 f^{abc}A_\mu^a \omega_\mu^{bc}\right)}_{***}\nonumber\\
    &=& (s+\delta)\int \d^dx\left(g f^{abc} \p_\mu \overline \omega_\nu^{ae}    D_\mu^{bd} c^d  \varphi_\nu^{ce}\right)-\delta\int \d^dx\left(g f^{abc} \p_\mu \overline \omega_\nu^{ae}    D_\mu^{bd} c^d  \varphi_\nu^{ce}\right)\,.
    \end{eqnarray}
We can rewrite this as
\begin{eqnarray}\label{GZb16}
(s+\delta) \tilde S_{\GZ}^{mod}&=& -\delta\int \d^dx\left(g f^{abc} \p_\mu \overline \omega_\nu^{ae}    D_\mu^{bd} c^d  \varphi_\nu^{ce}\right)=-\theta^2\int \d^dx\left(g f^{abc} \p_\mu \beta_\nu^{ae}    D_\mu^{bd} c^d  \varphi_\nu^{ce}\right)\,,
\end{eqnarray}
whereby we introduced a modified version of the original GZ action \eqref{GZb1}, given by $\tilde S_{\GZ}^{mod}=S_{\GZ}+ S_{aux}$. Looking at \eqref{GZb16}, we have found that $s+\delta$ ``almost'' generates a symmetry of the foregoing action $\tilde S_{\GZ}^{mod}$. In order to get an actual symmetry, we rewrite, using partial integration,
\begin{eqnarray*}\label{GZb16c}
  -\int \d^dx\left(g f^{abc} \p_\mu \beta_\nu^{ae}    D_\mu^{bd} c^d  \varphi_\nu^{ce}\right)&=&  \int \d^dxD_\mu^{bd}\left(g f^{abc} \p_\mu \beta_\nu^{ae}\varphi_\nu^{ce}  \right)[(\p_\nu D_\nu)^{-1}]^{dq}\frac{\delta S_{\GZ}^{mod}}{\delta \oc^q}\,.
\end{eqnarray*}
We can localize the latter term again analogously as before, by extending the auxiliary action by a novel quartet of fields,
\begin{eqnarray}
s Q^a ~=~  R^a\,, s R^a ~=~0 \,, s \overline R^a ~=~ \overline Q^a\,, s \overline Q^a ~=~0 \;.
\end{eqnarray}
with $R$, $\overline R$ anti-commutating fields, while $Q$ and $\overline Q$ are bosonic fields. We introduce a second auxiliary action
\begin{eqnarray}\label{GZb11}
S_{aux, 2}&=& s\int \d^dx\left(Q^a \p_\mu D^{ab}_\mu \overline R^b - \overline R^d \underbrace{D_\mu^{bd} (g f^{abc} \p_\mu \beta_\nu^{ae} \varphi^{ce}_\nu}_{\kappa^d })\right) \nonumber\\
&=& \int \d^dx\left( R^a \p_\mu D^{ab}_\mu \overline R^b +  Q^a \p_\mu D_\mu^{ab}\overline Q^b + g f^{abc}\p_\mu Q^a  D_\mu^{bd} c^d \overline R^c - \overline Q^d \kappa^d + \overline R^d s (\kappa^d) \right)
\end{eqnarray}
and further extend/adapt the $\delta$-transformation \eqref{GZb14} to
\begin{eqnarray}\label{GZb14b}
  \delta \overline Q^a &=&\theta^2 c^a\,, \quad   \delta \overline{c}^a~=~- \theta^2 Q^a-\theta^2 \alpha^a\,.
\end{eqnarray}
We finally introduce the following action
\begin{eqnarray}\label{GZmod}
S_{\GZ}^{mod}&=& \tilde S_{\GZ}^{mod}+S_{aux, 2}~=~S_{\GZ}+ S_{aux}+S_{aux, 2}\nonumber\\
&=&S_\YM +  \int \d^d x\,\left( b^a \p_\mu A_\mu^a +\overline c^a \p_\mu D_\mu^{ab} c^b \right)\nonumber  \\&&+ \int \d^d x\left( \overline \varphi_\mu^{ac} \p_\nu D_\nu^{ab} \varphi_\mu^{bc}  - \overline \omega_\mu^{ac} \p_\nu D_\nu^{ab} \omega_\mu^{bc}  +\theta^{2}  f^{abc}A_\mu^a \left( \varphi_\mu^{bc} +  \overline \varphi_\mu^{bc}\right)- g f^{abc} \p_\mu \overline \omega_\nu^{ae}    D_\mu^{bd} c^d  \varphi_\nu^{ce}\right)\nonumber\\
&&+\int \d^dx\left(\Omega^a \p_\mu D_\mu^{ab}\overline \Omega^b + \alpha^a \p_\mu D_\mu^{ab}\overline \alpha^b + gf^{abc}(\p_\mu \alpha^a)(D_\mu^{bd}c^d)\overline \Omega^c-\overline\alpha^a D_\nu^{ab} \overline\Lambda_\nu^b+\overline\Omega^a s(D_\nu^{ab} \overline\Lambda_\nu^b)\right.\nonumber\\&&
\left.+ \Psi_\nu^{ac}\p_\mu D_\mu^{ab}\overline \Psi_\nu^{bc}+\beta_\nu^{ac}\p_\mu D_\mu^{ab}\overline \beta_\nu^{bc} +gf^{abc} (\p_\mu\beta_{\nu}^{ae})(D_\mu^{bd}c^d)\overline \Psi_\nu^{ce}-f^{abc}A_\mu^a\overline{\beta}_\mu^{bc}-f^{abc}\overline{\Psi}_\mu^{bc}D_\mu^{ad}c^d\right)\nonumber\\
 &&+\int \d^d x\left( R^a \p_\mu D^{ab}_\mu \overline R^b +  Q^a \p_\mu D_\mu^{ab}\overline Q^b + g f^{abc}\p_\mu Q^a  D_\mu^{bd} c^d \overline R^c - \overline Q^d \kappa^d + \overline R^d s (\kappa^d) \right)
 \end{eqnarray}
and modified BRST transformation $s_\theta \equiv s + \delta$,
\begin{eqnarray}\label{GZb18}
s_\theta A_{\mu }^{a} &=&-\left( D_{\mu }c\right) ^{a}\,,\quad  s_\theta c^{a} ~=~\frac{1}{2}gf^{abc}c^{b}c^{c}\,,\quad s_\theta \overline{c}^{a} ~=~b^{a}-\theta^2\alpha^a -\theta^2 Q^a  \,,\quad  s_\theta b^{a}~=~0\,,  \nonumber \\
s_\theta \varphi _{\mu}^{ac} &=&\omega_{\mu}^{ac}\,,\quad s_\theta\omega_{\mu}^{ac}~=~0\,,\quad s_\theta \overline{\omega}_{\mu}^{ac} ~=~\overline{\varphi }_{\mu}^{ac}+\theta^2\beta_\mu^{bc}\,,\quad s_\theta \overline{\varphi }_{\mu}^{ac}~=~0\,,\nonumber\\
 s_\theta \alpha^a &=&\Omega^a\,, \quad  s_\theta\Omega^a ~=~0\,,\quad s_\theta \overline\Omega^a ~=~\overline\alpha^a\,, \quad  s_\theta\overline\alpha^a ~=~\theta^2 c^a\,,\nonumber\\
s_\theta \beta_\mu^{ab} &=&\Psi_\mu^{ab}\,,\quad   s_\theta\Psi_\mu^{ab}~=~0\,,\quad s_\theta \overline\Psi_\mu^{ab} ~=~\overline\beta_\mu^{ab}\,,\quad  s_\theta\overline\beta_\mu^{ab}~=~\theta^2 \omega_\mu^{ab}\,,\nonumber\\ s_\theta Q^a &=&  R^a  \,, \quad s_\theta R^a ~=~0 \,,\quad s_\theta \overline R^a ~=~ \overline Q^a\,,\quad s_\theta \overline Q^a ~=~\theta^2 c^a\,.
\end{eqnarray}
Combining all information, we consequently find
\begin{eqnarray}\label{GZmod4}
s_\theta S_{\GZ}^{mod}&=& (s+\delta)\left(\tilde S_{\GZ}^{mod}+S_{aux, 2}\right)\nonumber\\
&=& -\theta^2\int \d^dx\left(g f^{abc} \p_\mu \beta_\nu^{ae}    D_\mu^{bd} c^d  \varphi_\nu^{ce}\right)-   \theta^2\int \d^d x Q^a \p_\mu D_\mu^{ab} c^b+\int \d^d x (- \theta^2 c^d \kappa^d + \theta^2 Q^a \p_\mu D_\mu^{ab} c^b)\nonumber\\
&=&0
\end{eqnarray}
In summary, the modified GZ action $S_{\GZ}^{mod}$ enjoys a ``BRST" invariance generated by $s_\theta$. We notice that we did not have to use the equations of motion to identify this $s_\theta$-symmetry. The newly constructed BRST is not nilpotent, $ s_\theta^2\neq0$, in analogy with its nonlocal version written down in \eqref{GZb7}. We have that $s_\theta^4 = 0$. Upon using the equations of motion of the new fields, we can derive from \eqref{GZb18} the on-shell (but nonlocal) BRST invariance of the original GZ action. Notice that this BRST will be slightly different from the one constructed in \cite{Sorella:2009vt}, the difference being caused by the fact that we constructed a BRST invariance of the \emph{complete} GZ action \eqref{GZb1} rather than of \eqref{GZb4}.

\section{A few properties of the modified GZ action}
\subsection{Other symmetries}
We shall identify some extra invariances of the modified GZ action $S_{\GZ}^{mod}$. Firstly, as
\begin{eqnarray}\label{GZb19}
 \int \d^dx \left(\overline\Omega^a \frac{\delta}{\delta \alpha^a} -\overline\alpha^a \frac{\delta}{\delta \Omega^a}\right) S_{\GZ}^{mod} &=&\int \d^dx\left( \overline \Omega^a \p D^{ab} \overline \alpha^b - \overline \alpha^a \p D^{ab} \overline \Omega^b +  gf^{abc}(\p_\mu \overline\Omega^a)(D_\mu^{bd}c^d)\overline\Omega^c\right)
\end{eqnarray}
and by rewriting the r.h.s.~of it by means of
\begin{eqnarray}
\int \d^dx\left( \overline \Omega^a \p D^{ab} \overline \alpha^b - \overline \alpha^a \p D^{ab} \overline \Omega^b \right) &=& \int \d^dx\left( g f^{akb} \p A^k \overline \alpha^b \right) = \int \d^dx\left( g f^{akb}  \overline \alpha^b \frac{\delta}{\delta\overline b^k}S_{\GZ}^{mod}\right)\,,  \nonumber\\
\int \d^dx\left(-gf^{abc}(\p_\mu \overline\Omega^a)(D_\mu^{bd}c^d)\overline\Omega^c\right) &=& \int \d^dx\left(-\frac{1}{2}gf^{abc}(D_\mu^{bd}c^d)\p_\mu (\overline\Omega^a\overline\Omega^c)\right)\nonumber\\
       &=&\int \d^dx\left(\frac{1}{2}gf^{abc}\overline\Omega^a\overline\Omega^c\frac{\delta}{\delta\overline c^b}S_{\GZ}^{mod}\right)\,,
\end{eqnarray}
we conclude that
\begin{equation}\label{GZb20}
    \Delta_1= \int \d^dx\left(\overline\Omega^a \frac{\delta}{\delta \alpha^a}-\overline\alpha^a \frac{\delta}{\delta \Omega^a}-\frac{1}{2}gf^{abc}\overline\Omega^a\overline\Omega^c\frac{\delta}{\delta\overline c^b} - g f^{akb} \overline \Omega^a \overline \alpha^b \frac{\delta }{\delta b^k}\right)\,,\qquad \text{with }\Delta_1^2=0\,,
\end{equation}
establishes a (nilpotent) symmetry of $S_{\GZ}^{mod}$. Similarly, we also find the following symmetries of the action
\begin{equation}\label{GZb21}
\Delta_2 = \int \d^dx\left(\overline\Psi_{\nu}^{bc} \frac{\delta}{\delta \beta_\nu^{bc}}-\overline\beta_\nu^{bc} \frac{\delta}{\delta \Psi_\nu^{bc}} - \frac{1}{2}gf^{abc}\overline\Psi_\nu^{ae}\overline\Psi_\nu^{ce}\frac{\delta}{\delta\overline c^b} - g f^{akb} \overline \Psi^{ae}_\nu \overline \beta^{be}_\nu \frac{\delta }{\delta b^k} \right) \,,\qquad \text{with }\Delta_2^2=0\,,
\end{equation}
and
\begin{equation}
\Delta_3= \int \d^dx\left(\overline R^a \frac{\delta}{\delta Q^a}-\overline Q^a \frac{\delta}{\delta R^a}-\frac{1}{2}gf^{abc}\overline R^a \overline R^c\frac{\delta}{\delta\overline c^b} - g f^{akb} \overline R^a \overline Q^b \frac{\delta }{\delta b^k}\right)\,,\qquad \text{with }\Delta_3^2=0\,.
\end{equation}
We can also link some of the original fields with the newly introduced one through the symmetries
\begin{equation}
\Delta_4= \int \d^dx\left(\overline \Omega^a \frac{\delta}{\delta \overline c^a}+c^a \frac{\delta}{\delta \Omega^a} - g f^{akb} c^a \overline \Omega ^b \frac{\delta }{\delta b^k}\right)\,,\qquad \text{with }\Delta_4^2=0\,,
\end{equation}
and
\begin{equation}
\Delta_5= \int \d^dx\left(\overline R^a \frac{\delta}{\delta \overline c^a}+c^a \frac{\delta}{\delta R^a} - g f^{akb} c^a \overline R ^b \frac{\delta }{\delta b^k}\right)\,,\qquad \text{with }\Delta_5^2=0\,.
\end{equation}
Clearly, $\left\{\Delta_i,\Delta_j\right\}=0$, but $\left\{\Delta_i,s_\theta\right\}\neq0$ generate further symmetries. In addition, there might be more symmetries not related to this algebra, but we did not attempt to find such in this letter.

We notice that we can rewrite $S_{\GZ}^{mod}$ as
\begin{align}\label{GZb22}
S_{\GZ}^{mod}=&S_\YM +  \int \d^d x\,\left( b^a \p_\mu A_\mu^a +\overline c^a \p_\mu D_\mu^{ab} c^b \right) + \int \d^d x\left( \overline \varphi_\mu^{ac} \p_\nu D_\nu^{ab} \varphi_\mu^{bc}  - \overline \omega_\mu^{ac} \p_\nu D_\nu^{ab} \omega_\mu^{bc}  +\theta^{2}  f^{abc}A_\mu^a \left( \varphi_\mu^{bc} +  \overline \varphi_\mu^{bc}\right)\right)\nonumber\\
&\int \d^d x\left(-gf^{abc}\p_\mu \overline\omega_\nu^{ae}D_\mu^{bd}c^d\varphi_\nu^{ce}+\Omega^a \p_\mu D_\mu^{ab}\overline \Omega^b + \alpha^a \p_\mu D_\mu^{ab}\overline \alpha^b + gf^{abc}(\p_\mu \alpha^a)(D_\mu^{bd}c^d)\overline \Omega^c +\right.\nonumber\\
& + \Psi_\nu^{ac}\p_\mu D_\mu^{ab}\overline \Psi_\nu^{bc}+\beta_\nu^{ac}\p_\mu D_\mu^{ab}\overline \beta_\nu^{bc} +gf^{abc} (\p_\mu\beta_{\nu}^{ae})(D_\mu^{bd}c^d)\overline \Psi_\nu^{ce}  \nonumber\\
& \left.+ R^a \p_\mu D^{ab}_\mu \overline R^b +  Q^a \p_\mu D_\mu^{ab}\overline Q^b + g f^{abc}\p_\mu Q^a  D_\mu^{bd} c^d \overline R^c \right)+\Delta_1\int \d^d x \left(\Omega^a D_\nu^{ab} \Lambda_\nu^b+\alpha^a s(D_\nu^{ab} \Lambda_\nu^b)\right)\nonumber\\
& +\Delta_2\int \d^dx\left(f^{abc}A_\mu^a\Psi_\mu^{bc}-f^{abc}\beta_\mu^{bc}D_\mu^{ad}c^d\right) + \Delta_3 \int \d^d x \left(R^d \kappa^d +  Q^d s (\kappa^d))\right)\,.
\end{align}
Finally, we also observe that the action is left invariant under constant shifts of the fields $\Psi_\mu^{ac}$, $\overline c^a$, $\overline\omega_\mu^{ac}$, $\beta_\mu^{ac}$, $\alpha^a$, $\Omega^a$, $Q^a$ and $R^a$, expressed through the following identities
\begin{equation}\label{GZb22b}
    \int \d^d x \frac{\delta S_{GZ}^{mod}}{\delta \chi}=   0\,,\qquad \chi\in \left\{\Psi_\mu^{ac},\overline c^a,\overline\omega_\mu^{ac},\beta_\mu^{ac},\alpha^a,\Omega^a,Q^a,R^a\right\}\,.
\end{equation}

\subsection{Connection between the original Yang-Mills and modified GZ action in the case of vanishing Gribov mass}
An important property of the modified GZ action is to investigate what happens when we set $\theta^2=0$. If our $S_{\GZ}^{mod}$ is meaningful, we expect to find back the original Yang-Mills theory in the Landau gauge (modulo trivial, unity-related, terms in the action). Setting $\theta^2=0$ yields
\begin{eqnarray}\label{GZb26}
\left.S_{\GZ}^{mod}\right|_{\theta^2=0}   &=& S_\YM +  \int \d^d x\,\left( b^a \p_\mu A_\mu^a +\overline c^a \p_\mu D_\mu^{ab} c^b \right)  \nonumber\\
&&+ \int \d^d x\left( \overline \varphi_\mu^{ac} \p_\nu D_\nu^{ab} \varphi_\mu^{bc}  - \overline \omega_\mu^{ac} \p_\nu D_\nu^{ab} \omega_\mu^{bc}  -  \underline{g f^{abc} \p_\mu \overline \omega_\nu^{ae}    D_\mu^{bd} c^d  \varphi_\nu^{ce}}\right)\nonumber\\
&&+\int \d^d x\left(\Omega^a \p_\mu D_\mu^{ab}\overline \Omega^b + \alpha^a \p_\mu D_\mu^{ab}\overline \alpha^b +  \underline{gf^{abc}(\p_\mu \alpha^a)(D_\mu^{bd}c^d)\overline \Omega^c } \right.\nonumber\\&&
\left.+ \Psi_\nu^{ac}\p_\mu D_\mu^{ab}\overline \Psi_\nu^{bc}+\beta_\nu^{ac}\p_\mu D_\mu^{ab}\overline \beta_\nu^{bc} +  \underline{gf^{abc} (\p_\mu\beta_{\nu}^{ae})(D_\mu^{bd}c^d)\overline \Psi_\nu^{ce} } \right)\nonumber\\
 &&+\int \d^d x\left( R^a \p_\mu D^{ab}_\mu \overline R^b +  Q^a \p_\mu D_\mu^{ab}\overline Q^b +  \underline{ g f^{abc}\p_\mu Q^a  D_\mu^{bd} c^d \overline R^c} -  \underline{\overline Q^d \kappa^d }+  \underline{\overline R^d s (\kappa^d)} \right)\,,
\end{eqnarray}
with
\begin{equation}\label{GZb27}
    \kappa^d = D_\mu^{bd} (g f^{abc} \p_\mu \beta_\nu^{ae} \varphi^{ce}_\nu )\,.
\end{equation}
Considering (ghost neutral) Green functions of the original Yang-Mills fields, it is then easily checked that the underlined terms will never contribute, as there are no propagators of the desired kind to attach them to the Green functions of interest. This argument is completely similar to the one given originally in \cite{Zwanziger:1992qr} which was related to the presence of the term $g f^{abc} \p_\mu \overline \omega_\nu^{ae}    D_\mu^{bd} c^d  \varphi_\nu^{ce}$ in the GZ action, which can also not couple. The residual terms are all forming unities, and upon integrating out these, we recover the Yang-Mills action.  Glancing at \eqref{GZb18}, we notice that $s_{\theta=0}=s$, with $s$ the original BRST, upon generalization to the new fields which become pairwise BRST $s$-doublets. In fact, since $s$ is now a symmetry and seeing that
\begin{equation}\label{Gzb27b}
    \left.S_{\GZ}^{mod}\right|_{\theta^2=0}   = S_\YM +  \int \d^d x\,\left( b^a \p_\mu A_\mu^a +\overline c^a \p_\mu D_\mu^{ab} c^b \right)+s\,\int \d^d x \left(\ldots\right)\,,
\end{equation}
we did nothing more than writing down a somewhat more complicated version of the FP Landau gauge fixing. We shall thence also recover the original Yang-Mills BRST cohomology. All GZ-related fields, old or new, are then physically trivial as they appear as doublets of a nilpotent symmetry \cite{Piguet:1995er}.

\subsection{Connection between the modified and original GZ action}
In this section we wish to verify that $S_{\GZ}^{mod}$ and $S_{GZ}$ are in fact equivalent in the sense that for any Green function built from the original GZ fields, i.e.~ with $\phi\in\left\{A_\mu^a,b^a,\overline{c}^a,c^a,\varphi_\mu^{ab},\overline\varphi_\mu^{ab},\omega_\mu^{ab},\overline\omega_\mu^{ab}\right\}$, we have the following identification
\begin{eqnarray}\label{GZb32}
  \Braket{\phi(x_1)\ldots\phi(x_n)}_{mod}=  \int [\d\Phi]_{mod} \phi(x_1)\ldots\phi(x_n) e^{-S_{\GZ}^{mod}}=\int [\d\Phi]_{\GZ} \phi(x_1)\ldots\phi(x_n) e^{-S_{\GZ}}\,.
\end{eqnarray}
Let us show this in 2 ways. Firstly, the nonlocal substitutions
\begin{eqnarray}\label{GZb33}
\Omega^a &=& \Omega'^a - g f^{\ell b q} (\p_\mu \alpha^\ell) (D_\mu^{bd} c^d) [\p D^{-1}]^{qa} + s (D^{qk}_\nu \overline \Lambda^k_\nu ) [(\p D^{-1})]^{qa}\,,  \nonumber\\
\Psi_\nu^{ae} &=& {\Psi'}_\nu^{ae} -g f^{\ell b q} (\p_\mu \beta^{ae}_\nu) (D_\mu^{bd} c^d) [\p D^{-1}]^{qa} - f^{\ell q e} (D^{\ell d}_\nu c^d ) [(\p D^{-1})]^{qa}\,,\nonumber\\
R^a &=& R'^a - g f^{\ell b q} (\p_\mu Q^\ell) (D_\mu^{bd} c^d) [\p D^{-1}]^{qa} +s (\kappa^q  ) [(\p D^{-1})]^{qa} \,,\quad \alpha^a~=~\alpha'^a + D_\nu ^{qd} \overline \Lambda^d_\nu [\p D^{-1} ]^{qa}\,, \nonumber\\
Q^a &=& Q'^a +  \kappa^q   [\p D^{-1} ]^{qa}\,,\quad \beta_\nu^{ac}~=~{\beta'}_\nu^{ac}+ f^{dqc} A^d_\nu [ (\p D)^{-1}]^{qa}\,,
\end{eqnarray}
which come with a trivial Jacobian, lead to
\begin{eqnarray}\label{GZb34}
  \Braket{\phi(x_1)\ldots\phi(x_n)}_{mod}=  \int [\d\Phi]_{mod} \phi(x_1)\ldots\phi(x_n) e^{-{S'}_{\GZ}^{mod}}
\end{eqnarray}
after dropping the prime-notation again, with the shifted ${S'}_{\GZ}^{mod}$ given by
\begin{eqnarray}\label{GZb35}
{S'}_{\GZ}^{mod}&=&S_\YM +  \int \d^d x\,\left( b^a \p_\mu A_\mu^a +\overline c^a \p_\mu D_\mu^{ab} c^b \right)  \\
&&+ \int \d^d x\left( \overline \varphi_\mu^{ac} \p_\nu D_\nu^{ab} \varphi_\mu^{bc}  - \overline \omega_\mu^{ac} \p_\nu D_\nu^{ab} \omega_\mu^{bc}  +\theta^{2}  f^{abc}A_\mu^a \left( \varphi_\mu^{bc} +  \overline \varphi_\mu^{bc}\right) - g f^{abc} \p_\mu \overline \omega^{ae}_\nu D^{bd}_\mu c^d \varphi^{ce}_\nu\right)\nonumber\\
&&+\int \d^d x\left( \Omega^a \p_\mu D_\mu^{ab}\overline\Omega^b + \alpha^a \p_\mu D_\mu^{ab} \overline \alpha^b  + \Psi^{ac}_\nu \p_\mu D_\mu^{ab} \overline \Psi^{bc}_\nu + \overline{\beta}_\nu^{ac}\p_\mu D_\mu^{ab}\beta_\nu^{bc} + R^{a} \p_\mu D_\mu^{ab}R^{b} +  Q^{a} \p_\mu D_\mu^{ab} Q^{b}\right)\nonumber\,.
\end{eqnarray}
Consequently, we can perform the path integration over the new fields, which are pairwise unities, to discover that
\begin{eqnarray}\label{GZb36}
\Braket{\phi(x_1)\ldots\phi(x_n)}_{mod}=  \int [\d\Phi]_{\GZ} \phi(x_1)\ldots\phi(x_n) \e^{-S_{\GZ}}=\Braket{\phi(x_1)\ldots\phi(x_n)}_{\GZ}\,.
\end{eqnarray}
This important formula means that the original GZ correlation functions can be evaluated with either the original or the modified GZ action. We may thus replace the original GZ action with nonlocal BRST with its modified version, enjoying a local version of the BRST. As a first corollary, we can reestablish that $\left.S_{\GZ}^{mod}\right|_{\theta^2=0}$ is equivalent with normal Yang-Mills gauge theories, as we know that GZ is for $\theta^2=0$.

Secondly, to avoid the use of nonlocal shifts to prove the important result \eqref{GZb36}, let us also present an alternative derivation. We reconsider the generic correlation function \eqref{GZb32}
and consequently recall the following Gaussian integration formula in case of two cc fields,
\begin{equation}\label{GZb50}
    \int \d\sigma d\overline\sigma \e^{-\int \d^d x~ \left(\overline\sigma ~\Delta~ \sigma + J~\overline\sigma+\overline J~ \sigma\right)}\propto \det \Delta^{-1} \e^{-\overline J~ \Delta~ J}\,,
\end{equation}
which means that if $J=0$ or $\overline J=0$, we shall just pick up a determinant upon integration. A similar formula holds for anti-commuting fields, yielding the inverse power of the determinant. If we then apply this to \eqref{GZb32} and adopt the integration
order $(\overline R, R)$, $(\overline Q, Q)$, $(\overline \Psi, \Psi)$, $(\overline \beta, \beta)$, $(\overline \Omega, \Omega)$, $(\overline \alpha, \alpha)$,  it is easily seen that everything neatly cancels, including the determinants. This once again leads to the result \eqref{GZb36}. As a second corollary, we can mention that the gap formulation of the horizon condition, $\frac{\p \Gamma}{\p \gamma^2}=0$, will also remain unchanged.

\sect{From GZ to RGZ}
So far, we have handled the original Gribov-Zwanziger theory. We recall that we have introduced in \cite{Dudal:2007cw,Dudal:2008sp} the Refined Gribov-Zwanziger (RGZ) theory, in which additional nonperturbative effects resulted in a dynamical mass scale in the $\{\overline{\varphi}_\mu^{ac},\varphi_\mu^{ac},\overline\omega_\mu^{ac},\omega_\mu^{ac}\}$ sector, see also \cite{Gracey:2010cg}. This extra scale seems to be necessary to accommodate for a nonvanishing infrared gluon and non-enhanced infrared ghost propagator in the GZ framework, in consistency with Landau gauge lattice data known up to date\footnote{There has been a lattice inspired proposal that different results for the propagators in the infrared would correspond to different nonperturbative extensions of the Landau gauge \cite{Maas:2009se}. Further studies are required to test this proposal, see also \cite{Blank:2010pa}.}, see e.g.~\cite{Cucchieri:2007md,Bogolubsky:2009dc,Dudal:2010tf,Bornyakov:2009ug}.

One might wonder if we could introduce the relevant RGZ operator $\overline{\varphi}_\mu^{ac}\varphi_\mu^{ac}-\overline\omega_\mu^{ac}\omega_\mu^{ac}$ without spoiling the $s_\theta$ BRST invariance.  We recall that this operator is BRST $s$-exact, so in the original GZ formulation a nonvanishing VEV $\Braket{\overline{\varphi}_\mu^{ac}\varphi_\mu^{ac}-\overline\omega_\mu^{ac}\omega_\mu^{ac}}\neq0$ was certainly possible due to the already (softly) broken BRST $s$. Fortunately, it appears to be possible to maintain the in this paper constructed local BRST invariance $s_\theta$,  even after ``refining'' the modified GZ theory\footnote{Which is still necessary as we would like to maintain the described gluon and ghost propagator behaviour.}. Of course, a in depth treatment would need the full study of the renormalizability, amongst other things, but in this letter we suffice by noticing that
\begin{equation}\label{GZb31c}
s_\theta \mathcal{Q}_{RGZ}~=~0,\qquad  \Delta_{1,2,3,4,5}\mathcal{Q}_{RGZ}~=~0\qquad \text{for}\; \mathcal{Q}_{RGZ}~=~\overline{\varphi}_\mu^{ac}\varphi_\mu^{ac}-\overline\omega_\mu^{ac}\omega_\mu^{ac}+\overline{\beta}_\mu^{ac}\beta_\mu^{ac}-\overline\Psi_\mu^{ac}\Psi_\mu^{ac}\,.
\end{equation}
It is also readily verified that the identification \eqref{GZb36} remains valid after the introduction into the action of the $d=2$ operator $\mathcal{Q}_{RGZ}$. The shift invariances \eqref{GZb22} will be linearly broken.

\sect{Future applicability of the new BRST symmetry}
In this letter, we took a small step in the ambitious program outlined in the introduction concerning the study of glueballs in the (R)GZ setting. Here, we just constructed a symmetry that \emph{might} be useful in order to construct a would-be physical subspace. We already mentioned that $s_\theta$ is not nilpotent, as such we loose the cohomology toolbox. We should stress here that our new BRST transformation $s_\theta$ is also particularly constructed in the Landau gauge, to which the GZ action \eqref{GZb1} applies to. As such, it has of course not the power of the original BRST transformation $s$ in ordinary perturbative gauge theories, where it is equivalent with local gauge invariance. Nonetheless, $s_\theta$ is an extension of the original transformation $s$ to the GZ case, and it is a nontrivial invariance, thus ideally suited to constrain the total Hilbert space of operators. We would like to point out that insisting on the nilpotency of $s_\theta$ would perhaps be asking too much. As explained in e.g.~\cite{Weinberg:1996kr}, the nilpotency of the usual BRST variation $s$ is necessary to be able to ``walk around'' freely in the space of Yang-Mills theories perturbatively fixed in a gauge of choice, and this without changing the physics. However, we cannot expect that we can still walk around freely, since the treatment of Gribov copies (\`{a} la Gribov-Zwanziger or any other way for that matters) depends on the gauge we are choosing. In particular, the construction of the Gribov-Zwanziger action relies on specific properties of the Landau gauge. This is a bit similar as to what happens in the Curci-Ferrari model specified to the Landau gauge \cite{Curci:1976bt}. Also there a not nilpotent BRST symmetry can be written down, see also the recent paper \cite{Tissier:2010ts,vonSmekal:2008en} for a discussion.

The lack of nilpotency of $s_\theta$ needs not to be a serious drawback in discussing renormalization aspects. Introducing the operators $s_1=s_\theta$, $s_2=s_\theta^2$, $s_3=s_\theta^3$, these 3 operators $s_i$ form a closed algebra. Following e.g.~\cite{Maggiore:1992ug,Fucito:1997xm}, it is then always possible, by the introduction of suitable global ghosts\footnote{The symmetry algebra can even be supplemented by other symmetry generators, if present.}, to construct a generalized nilpotent ``BRST'' operator $\mathcal{S}$, with $\mathcal{S}^2=0$, whose symmetry content is equivalent to that of the operators $s_i$. In general, the operators $s_i$ will obey an algebra
\begin{equation}\label{np1}
    \left[s_i,s_j\right]_{\pm}=f_{ijk}s_k\,,
\end{equation}
where the appropriate bracket is considered for (anti)-commuting operators. In our case, all $f_{ijk}$'s are actually zero, but let us keep them for generality. One then introduces global ghosts $\epsilon_{i}$ (again with appropriate (anti)-commuting character), to define
\begin{equation}\label{np2}
    \mathcal{S}=\epsilon_i s_i -\frac{1}{2}f_{ijk}\epsilon_i\epsilon_j\frac{\p}{\p \epsilon_k}\,.
\end{equation}
Evidently, $\mathcal{S}$ is a symmetry if, and only if, the $\delta_i$'s are, since the action does not depend on the global ghosts. By construction, one finds
\begin{equation}\label{np3}
    \mathcal{S}^2=0\,.
\end{equation}
The quantum version of $\mathcal{S}$ can then be used to investigate the stability of the action using the associated, nilpotent, Slavnov-Taylor identity, as explained in \cite{Fucito:1997xm}. However, $\mathcal{S}$ does not help at all in the selection of a physical subspace, in the sense that states annihilated by the charge $\mathcal{Q}_\mathcal{S}$, modulo the exact ones, will contain no more (or less) information than those annihilated by the $\mathcal{Q}_{s_i}$ charges. This is obvious, since the symmetry constraints set by the $s_i$ are identical to those set by $\mathcal{S}$. It is perhaps interesting to refer here also to the BRST charge $s_{m}$ of the Curci-Ferrari model, where also $s_m^2\neq0$. Using the conventions and notations of \cite{Delduc:1989uc}, it is apparent that $s_m^4=0$. So, in principle, also for the Curci-Ferrari case one can built a nilpotent BRST operator $\mathcal{S}$, which evidently cannot guarantee the unitarity of the Curci-Ferrari model, but which is nevertheless a powerful tool in discussing renormalization effects.

In the usual (perturbative) Yang-Mills context, the algebra of the nilpotent BRST operator $s$ and Faddeev-Popov ghost charge is used to classify representations. We would however like to point out that the mere physicalness of the subspace defined by that BRST charge is not ensured by the BRST symmetry itself, nor its cohomology. Indeed, the positivity, a crucial ingredient for a unitary physical $S$-matrix, is checked explicitly, independent of the BRST. In particular, in the perturbative Faddeev-Popov context well known from textbooks, the two remaining degrees of freedom are the massless transverse polarizations, and these do indeed display a positive measure, which is verified after the appropriate commutation relations are introduced in the common way. The fact that the measure of the remaining would-be physical degrees of freedom need to be checked explicitly, via the (anti)commutation relations, is mentioned in eq.~(3.6) and the surrounding discussion in the classical reference paper \cite{Kugo:1979gm}; see also Chapter 14 in \cite{Henneaux:1992ig}. We thus strongly point out here that the concept of a BRST cohomology\footnote{Be it of the original BRST in perturbative Yang-Mills theories, or in any other setting with a BRST symmetry, including the ``large BRST'' operator $\mathcal{S}$ mentioned in the previous paragraph.} is thus never sufficient on its own to ensure a physical subspace. It should be clear that the assumption of a positive measure on the remaining subspace is a difficult one to check when discussing composite operators, especially when one tries to get an understanding of the nonperturbative sector of gauge theories, where confinement sets in. The foregoing comments do not only apply to the GZ case studied here, but also to studies of propagators in e.g.~the Schwinger-Dyson approach. Simply claiming that the BRST, if even present to begin with, is sufficient for a physical subspace of confined colorless particles is only part of the story. To our knowledge, nobody ever presented an explicit check in the Schwinger-Dyson context (or any other for that matters) that composite operators belonging to the BRST cohomology do display positivity.

\subsection{The noninteracting level}
As an example, let us briefly look at the construction of the scalar glueball operator. Our first concern is at the noninteracting (= quadratic) level. In that case, we notice that the $\Delta_{1,2,3}^{nonint}$ symmetries handle the new fields as doublets. This means that, at the quadratic level, an operator $\mathcal{O}$, containing\footnote{In the current quadratic approximation, we only need to refer to the Abelian part $f_{\mu\nu}^2$ of the complete field strength $F_{\mu\nu}^2$.} $f_{\mu\nu}^2$, consistent with $s_\theta \mathcal{O}=\Delta_{1,2,3}\mathcal{O}=0$ can only be of the form
\begin{equation}\label{GZn1}
    \mathcal{O}=f_{\mu\nu}^2+\mathcal{O}_1(\text{original GZ fields})+\Delta_{1}^{nonint}(\ldots)+\Delta_{2}^{nonint}(\ldots)+\Delta_{3}^{nonint}(\ldots)\,.
\end{equation}
At the level of the correlation function, we shall thus find at lowest order
\begin{equation}\label{GZn2}
    \Braket{\mathcal{O} \mathcal{O}}_{mod}=\Braket{\left[f_{\mu\nu}^2+\mathcal{O}_1(\text{original GZ fields})\right]\left[f_{\mu\nu}^2+\mathcal{O}_1(\text{original GZ fields})\right]}_{mod}
\end{equation}
since $\Delta_{1,2,3,4,5}^{nonint}$ are symmetries of the quadratic action. The operators appearing in the r.h.s.~of \eqref{GZn2} are no longer depending on the new fields, thus we can apply the relation \eqref{GZb36} to replace the modified GZ action with the original one. As such, the quest for a physical glueball correlator at lowest order is reduced to the original question in the original GZ theory. The point is that this operator now obeys a certain well-certified symmetry constraint. This was missing in the original formulation, due to the lack of a local BRST-like invariance.

In particular, we can consider the following example,
\begin{equation}
O_{phys}^{GZ}=\frac{1}{2}f_{\mu\nu}^2+N s( \varphi_{\mu\nu}^a \overline \omega_{\mu\nu}^a)\,,
\end{equation}
where we introduced the general notation
\begin{equation}
    \rho_{\mu\nu}^a = \frac{1}{N} f^{abc}  ( \p_\mu \rho^{bc}_\nu - \p_\nu \rho^{bc}_\mu)\,.
\end{equation}
The analyticity structure of the two-point functions created with $O_{phys}^{GZ}$ is currently under study elsewhere using the setup of \cite{Baulieu:2009ha}. Since
\begin{eqnarray}
s_\theta O_{phys}^{GZ}&=&-\theta^2\omega_{\mu\nu}^a\beta_{\mu\nu}^a
\end{eqnarray}
and
\begin{equation}
    s_\theta\left[\overline\beta_{\mu\nu}^a\beta_{\mu\nu}^a-\overline\Psi_{\mu\nu}^a\Psi_{\mu\nu}^a\right]=\theta^2\omega_{\mu\nu}^a\beta_{\mu\nu}^a\,,
\end{equation}
we can introduce
\begin{equation}\label{ff}
O_{phys}^{mod}=O_{phys}^{GZ}+\overline\beta_{\mu\nu}^a\beta_{\mu\nu}^a-\overline\Psi_{\mu\nu}^a\Psi_{\mu\nu}^a\,,\qquad \text{with } s_\theta O_{phys}^{mod}=0\,.
\end{equation}
We can rewrite as follows:
\begin{equation}\label{ff2}
O_{phys}^{mod}=O_{phys}^{GZ}-\Delta_2\left[\beta_{\mu\nu}^a\Psi_{\mu\nu}^a\right]\,,
\end{equation}
so that
\begin{equation}\label{ff3}
\Braket{O_{phys}^{mod}(x)O_{phys}^{mod}(y)}_{mod}=\Braket{O_{phys}^{GZ}(x)O_{phys}^{GZ}(y)}_{GZ}\,,
\end{equation}
by using \eqref{GZb32}. At least at the noninteracting level, we would then conclude that no more physics is lurking in the modified GZ than in the original GZ theory. More precisely, we can construct an operator in the modified GZ consistent with the BRST $s_\theta$-symmetry constraint, and its expectation value can be computed using the original GZ action. It would of course remain to show what happens with this operator at the quantum (interacting) level, and to see if it actually has a decent analytical structure.

\subsection{The interacting level}
Turning now to the interacting level, an obstacle might arise under the form of the new fields being no more doublets under the new symmetries $\Delta_{1,2,3}$. For example, specifying to $\Delta_1$, we see that $\overline\alpha$ and/or $\overline{\Omega}$ also appear in the variation of $b$ and $\overline c$. Consequently, the following is no longer true:
\begin{equation}\label{Gzn4}
    \Delta_1\mathcal{G}=0\Rightarrow \mathcal{G}=\Delta_1(\ldots)\,,
\end{equation}
a simple counterexample being
\begin{equation}\label{GZn3}
    \mathcal{G}=\frac{1}{2}gf^{abc}\overline\Omega^a\alpha^c+\overline c^b\,.
\end{equation}
In general, the construction of the most general operator $\mathcal{O}$ consistent with (the linearized quantum extension of)
\begin{equation}\label{Gn3b}
    s_\theta \mathcal{O}~=~ \Delta_{1,2,3,4,5}\mathcal{O}~=~0
\end{equation}
can be quite complicated, since the BRST $s_\theta$ is not nilpotent, while the other symmetries $\Delta_{1,2,3,4,5}$ contain nonlinear variations. In addition the doublet theorem \cite{Piguet:1995er} does not apply anymore.

Nevertheless, we observe that the following operator
\begin{equation}\label{oop}
O_{phys}^{mod}=\frac{1}{2}F_{\mu\nu}^2+\omega_{\mu\nu}^a  \overline \omega_{\mu\nu}^a -  \varphi_{\mu\nu}^a  \overline \varphi_{\mu\nu}^a+\overline\beta_{\mu\nu}^a\beta_{\mu\nu}^a-\overline\Psi_{\mu\nu}^a\Psi_{\mu\nu}^a\,,
\end{equation}
which is a possible natural extension of \eqref{ff} to the interacting level, has the property to contain $F_{\mu\nu}^2$, while
\begin{equation}
s_\theta O_{phys}^{mod}=0\,,
\end{equation}
next to
\begin{equation}\label{ff2bis}
O_{phys}^{mod}=  \frac{1}{2}F_{\mu\nu}^2+\omega_{\mu\nu}^a  \overline \omega_{\mu\nu}^a -  \varphi_{\mu\nu}^a  \overline \varphi_{\mu\nu}^a  -\Delta_2\left[\beta_{\mu\nu}^a\Psi_{\mu\nu}^a\right]= O_{phys}^{GZ}-\Delta_2\left[\beta_{\mu\nu}^a\Psi_{\mu\nu}^a\right]\,,
\end{equation}
so that we would again find \eqref{ff3} and reach the some conclusion as written down thereafter, but this time using the full, interacting, theory.

Of course, we cannot make more definite statements at this point. A full study of the renormalization of the modified GZ action would be needed, as well as a study of the renormalization properties of the operator \eqref{oop}. Either one starts with $\frac{1}{2}F_{\mu\nu}^2$ and $\omega_{\mu\nu}^a  \overline \omega_{\mu\nu}^a -  \varphi_{\mu\nu}^a  \overline \varphi_{\mu\nu}^a+\overline\beta_{\mu\nu}^a\beta_{\mu\nu}^a-\overline\Psi_{\mu\nu}^a\Psi_{\mu\nu}^a$ as separate operators and after the renormalization analysis, find out whether they can be combined into a pure or at least less complicatedly mixed operator. Or, one starts the renormalization analysis directly with $O_{phys}^{mod}$ and see what happens with it at the quantum level.

The main goal of this letter was to show that one can replace the GZ partition function with a modified version of it, and this without changing the relevant physics of the GZ theory. Although the replacement comes at the cost of extra fields, the gain lies in the presence of a local, albeit not nilpotent, BRST symmetry, which is a natural extension of the usual BRST of ordinary Yang-Mills gauge theories. The main results of this letter are summarized by the action \eqref{GZmod}, the BRST symmetry \eqref{GZb18} and the identification relation \eqref{GZb36}.

We end with the comment that only a combined effort of a renormalization analysis (to control the quantum effects) and of the spectral representation (to control the physical relevance of the operator) can lead to definite statements about the physical meaning of a composite operator describing bound states, in particular the glueballs we wish to explore in the (R)GZ context. Other exploratory steps in this complicated task were set in \cite{Sorella:2010fs,Sorella:2010it,Dudal:2009zh,Baulieu:2009ha,Dudal:2010cd}.

\section*{Acknowledgments}
D.~Dudal and N.~Vandersickel are supported by the Research-Foundation
Flanders (FWO Vlaanderen).  We wish to thank S.~P.~Sorella and M.~Q.~Huber for useful feedback.

\end{document}